\documentclass[usenatbib]{mn2e}
\usepackage{graphicx,times}
\usepackage{natbib}
\usepackage{amssymb}
\usepackage{amsmath,bm}
\newcommand{\be}{\begin{equation}}
\newcommand{\ee}{\end{equation}}
\newcommand{\ba}{\begin{eqnarray}}
\newcommand{\ea}{\end{eqnarray}}
\newcommand{\no}{\nonumber}
\newcommand{\bi}{\begin{itemize}}
\newcommand{\ei}{\end{itemize}}
\newcommand{\bfi}{\begin{figure}
\epsfxsize=9cm \epsffile}
\newcommand{\efi}{\end{figure}}

\newcommand{\mnras}{MNRAS}
\newcommand{\apj}{ApJ}
\newcommand{\apjl}{ApJL}

\newcommand{\prd}{PRD}
\newcommand{\physrep}{Physics Reports}
\newcommand{\araa}{Annual Review of Astronomy \& Astrophysics}

\title[Improved lensing reconstruction]{Weak lensing
  reconstruction through cosmic magnification. II:  Improved
  power spectrum determination and map-making}
\author[  Yang et al.]{  Xinjuan
Yang$^{1}$, Pengjie
Zhang$^{2,1}$\thanks{Email: zhangpj@sjtu.edu.cn}, Jun Zhang$^{2}$, and YuYu$^{1}$\\
$^{1}$Key Laboratory for Research in Galaxies and Cosmology,
Shanghai Astronomical Observatory, Nandan Road 80, Shanghai, 200030,
China. \\
$^{2}$Center for Astronomy and Astrophysics, Department of
  Physics and Astronomy, Shanghai Jiao Tong
  University, 955 Jianchuan road, Shanghai, 200240}
\begin{document}
\maketitle
\begin{abstract}
The existence of galaxy intrinsic clustering severely hampers  the
weak lensing reconstruction from cosmic magnification. In paper I
\citep{Yang2011}, we proposed a minimal variance estimator to
overcome this problem. By utilizing the different dependences of
cosmic magnification and galaxy intrinsic clustering on galaxy flux,
we demonstrated that the otherwise overwhelming galaxy intrinsic
clustering can be significantly suppressed such that lensing maps
can be reconstructed with promising accuracy.  This procedure relies
heavily on the accuracy of determining the galaxy bias from the same
data. Paper I adopts an iterative approach, which degrades toward
high redshift. The current paper presents an alternative method,
improving over paper I. We prove that the measured galaxy clustering
between flux bins allows for simultaneous determination of the
lensing power spectrum and the flux dependence of galaxy bias, at
this redshift bin. Comparing to paper I, the new approach is not
only more straightforward, but also more robust. It identifies an
ambiguity in determining the galaxy bias and further discovers a
mathematically robust way to suppress this ambiguity to
non-negligible level ($\sim 0.1\%$). The accurately determined
galaxy bias can then be applied to the minimal variance estimator
proposed in paper I to  improve the lensing map-making. The gain at
high redshift is significant.    These maps can be used to measure
other statistics,  such as cluster finding and peak statistics.
Furthermore, by including galaxy clustering measurement between
different redshift bins, we can also determine the lensing cross
power spectrum between these bins, up to a small and correctable
multiplicative factor.
\end{abstract}
\begin{keywords}
cosmology: theory -- cosmological parameters -- gravitational
lensing -- dark matter
\end{keywords}

\section{introduction}
Cosmic magnification
\citep{Gunn1967,Blandford1992,Bartelmann1995,Dolag1997,Hamana2001,Menard2002a,
 Menard2002b,Menard2003b}, the lensing induced coherent fluctuation
in galaxy number distribution, offers an attractive alternative to
cosmic shear ( for reviews see
\citealt{Bartelmann2001,Schneider2006,Hoekstra2008,Munshi2008}) to
reconstruct the matter distribution of the universe. (1) It does not
require galaxy shape measurement and hence avoids all potential
problems associated with it. (2) It is even insensitive to
photometry errors. This is quite surprising, given that weak lensing
reconstruction through cosmic magnification indeed requires
galaxy/quasar flux measurement. This point will be further explained
in the appendix.

A formidable task in weak lensing reconstruction through cosmic
magnification is to reduce contamination caused by the galaxy
intrinsic clustering, which is in general overwhelming.
\cite{Zhang2005} argued that such contamination can be removed by
the distinctive flux dependences of the cosmic magnification signal
and the intrinsic clustering noise.  In a companion paper
(\cite{Yang2011}, hereafter paper I), we showed that such separation
is indeed doable. We constructed a minimal variance estimator for
the weak lensing map reconstruction.  It not only extracts the
lensing signal from the observed galaxy number distribution, but
also removes intrinsic galaxy clustering and minimizes the shot
noise simultaneously. Weak lensing maps can then be reconstructed
for each source redshift bin, through which one can measure the
lensing auto and cross power spectra. This reconstruction requires
no prior knowledge on the galaxy bias, other than that the
stochasticity between galaxy number density distributions of
different fluxes is not overwhelming.

Nevertheless, we noticed in paper I that the reconstruction accuracy
degrades at high redshift. It is hence worthwhile to explore new
approaches. The current paper proposes a promising alternative. It
is a two-step procedure. Firstly, we start with the measured galaxy
angular power spectra between different flux bins (but of the same
redshift bin). These are heavily reduced data comparing to the
``raw'' maps of galaxy number density distribution on the sky and
are much easier to analyze than the ``raw'' maps. They are the
mixtures of the galaxy intrinsic clustering (power spectra), the
lensing power spectrum and cross terms. We prove that, due to the
different dependence of cosmic magnification and galaxy intrinsic
clustering on galaxy flux, we can separate these components and
solve for the lensing power spectrum. This improves on paper I,
especially at high redshift.

The galaxy bias can also be determined simultaneously, with
significantly improved accuracy at high redshift. In particular we
find a degeneracy in determining the galaxy bias. This degeneracy
likely degrades the weak lensing map reconstruction at high
redshifts in paper I (Fig. 6). Fortunately, now we  find a
mathematically robust remedy to minimize its impact to $0.1\%$ on
the determined galaxy bias. This allows us to construct lensing maps
with the minimal variance estimator proposed in paper I.  Due to the
improved galaxy bias determination, the quality of maps is improved,
especially at high redshift. This is the second step. These maps can
be used for cluster finding, peak statistics and other non-Gaussian
statistics. In particular, they are useful for cross correlating
other cosmic fields such as CMB lensing
\citep{Seljak1999,Hu2002,Hirata2003,Smidt2011,Das2011,van
Engelen2012,Bleem2012, Das2013,Ade2013}, 21cm background lensing
\citep{Cooray2004,Pen2004,Zahn2006,Mandel2006}, cosmic shear
\citep{Waerbeke2000,Bacon2000,Hoekstra2002,Hoekstra2006,Massey2007,Fu2008,Lin2012,Jee2013},
galaxy distribution
 \citep{Kaiser1992,Menard2003a,Jain2003,Scranton2005,Zhang06,Hildebrandt2009,Waerbeke2010,Menard2010,Hildebrandt2011,
Ford2012}, the thermal Sunyaev Zel'dovich effect (e.g. the thermal
SZ tomography, \citealt{Shao2011}), the integrated Sachs-Wolfe
effect (\citealt{Loverde2007} and references therein ), and other
cosmic fields.

The paper is organized as follows.  In \S \ref{sec:method}, we
present our method to directly determine the lensing auto power
spectrum and galaxy bias. We make a performance of this approach to
SKA. We also find that the lensing cross power spectrum between
different redshift bins can be determined with the determined galaxy
bias previously. In \S \ref{sec:improvement}, we show how the
improved galaxy bias from the direct power spectrum determination
approach is to significantly improve the $\kappa$ map-making (paper
I). We discuss and summarize in \S \ref{sec:conclusions}. In
appendix,  we prove the uniqueness of the direct power spectrum
determination (appendix \ref{sec:appendix1}) and discuss why the
reconstruction from cosmic magnification is insensitive to the
photometry errors (appendix \ref{sec:appendix2}). The adopted
specifications of SKA and the fiducial model are the same as paper
I.

\section{Direct determination of lensing power spectrum through galaxy
 power spectrum measurements}
\label{sec:method}  For any given redshift bin, we can further split
galaxies into different flux bins and measure the galaxy number
density correlations between these flux bins. Since the galaxy
intrinsic clustering and cosmic magnification depend on the galaxy
luminosity in different ways, naively we expect that it  is possible
to measure the two simultaneously by directly fitting the measured
correlations. However, further investigation presented in this
section found a degeneracy between the intrinsic clustering and
cosmic magnification. Fortunately we found a simple but efficient
remedy.  It is able to render this degeneracy irrelevant for
realistic cases and enables  direct determination of the lensing
power spectrum feasible. In this section, \S
\ref{subsec:reconstruction}, \S \ref{subsec:error} and \S
\ref{subsec:ska} focus on correlations/power spectra within the same
redshift bin. In \S \ref{subsec:direct_cross}, we will discuss the
extension to cross correlation between different redshift bins.

\begin{figure}
\includegraphics[angle=270,width=84mm]{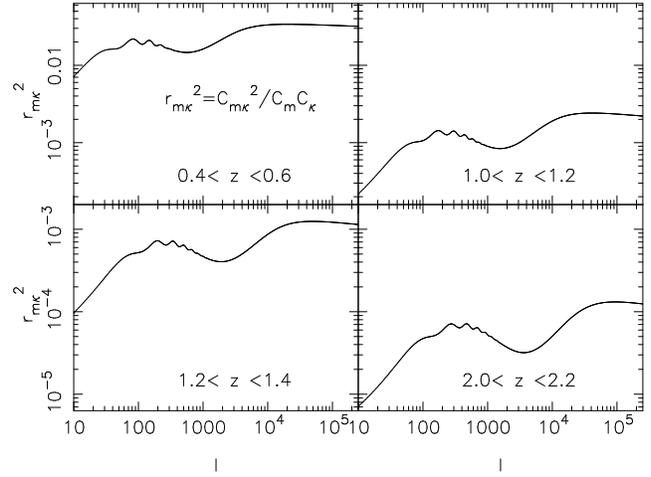}
\caption{Directly determined $\tilde{C}_\kappa$ differs from the
  true lensing power spectrum $C_{\rm \kappa}$  by a multiplicative factor $1-r^2_{\rm
    m\kappa}$. Here, $r_{\rm m\kappa}$ is the cross correlation
  coefficient between the matter distribution and convergence, of the given redshift
  bin. Due to the lensing kernel, $r^2_{\rm m\kappa}\ll 1$ and the
  induced bias is negligibly small, of the order $10^{-2}$-$10^{-4}$.
  Oscillations in the figure are caused by the Baryon
Acoustic Oscillations (BAO), so the oscillations move to larger
multipole with  increasing redshift. } \label{fig:one}
\end{figure}

\subsection{Direct determination of lensing auto power spectrum}
\label{subsec:reconstruction}

 For convenience, we will work in
Fourier space. For a flux limited survey, we divide galaxies into
flux and redshift bins. Throughout the paper, we use subscript
``$i$'' and ``$j$'' to denote the flux bins. For a given redshift
bin and a given scale $\bm{\ell}$, $\delta^{\rm L}_{{\rm
g},i}(\bm\ell)$ is the Fourier transform of the observed galaxy
over-density of the $i$-th flux bin. For brevity, we simply denote
it as $\delta_i^{\rm L}$ hereafter. With the observed galaxy power
spectra between all flux bins we are able to perform a direct weak
lensing power spectrum determination without any priors on the
galaxy bias except that it is deterministic.

The cosmic magnification effect changes the galaxy number
over-density to $\delta^{\rm L}_{i}=\delta_{i}+g_i\kappa$. Here
$\delta_{i}$ is the galaxy intrinsic clustering,
$g_i\equiv2(\alpha_i-1)$ and $\alpha_i$ is defined as the negative
logarithmic slope of the differential luminosity function minus one
\citep{Bartelmann1995,Broadhurst1995,Scranton2005,Zhang2005,Yang2011},
and $\kappa$ is the convergence \citep{Jain1997,Bartelmann2001}. It
is worth noting that, $\delta^{\rm L}_{i}$ and $g_i$ are measurable
quantities, but $\delta_i$ and $\kappa$ are unknowns. The observed
galaxy power spectrum between the $i$-th and $j$-th flux bins (but
of the same redshift bin) is \ba \label{eqn:measurements}
 \bar{C}_{ij}(\ell)&=&\left\langle\delta_{i}^{\rm L}({\bm
   \ell})\delta_{j}^{\rm L}(-{\bm \ell})\right\rangle \\
\nonumber &=&b_{i}b_{j}C_{\rm m}(\ell)+g_{i}g_{j}C_{\rm
\kappa}(\ell)+(g_{i}b_{j}+g_{j}b_{i})C_{\rm m\kappa}(\ell)\ .
 \ea
Here   $C_{\rm m}$, $C_{\rm m\kappa}$ and $C_\kappa$ are the matter
power spectrum, matter-lensing cross power spectrum
 and lensing power spectrum, respectively. They are all unknowns.
 $C_\kappa$ is the signal that we want to directly reconstruct. The
 unknown galaxy bias $b$ and the associated intrinsic clustering
 $b_ib_iC_m$ are the major uncertainties that we want to remove. This expression also assumes a deterministic
galaxy bias defined as $\delta_i=b_i\delta_{\rm m}$. $\delta_{\rm
m}$ is the dark matter surface over-density projected over the given
redshift bin with the same weighting as the galaxy number
distribution.

In the above  expression,  we have subtracted the ensemble
  averaged shot noise in the galaxy power spectrum measurement.
Nevertheless, statistical fluctuations in shot noise inevitably
introduce error in the power spectrum measurement. On the other
hand, the galaxy distribution does not totally trace the dark matter
distribution and the galaxy stochasticity exists
\citep{Pen1998,Seljak2009,Hamaus2010,Sato2013,Baldauf2013}. This
systematic galaxy stochasticity will introduce errors in the direct
power spectrum determination. We postpone the discussion until \S
\ref{subsec:error}.

Notice that in Eq. \ref{eqn:measurements},
$\bar{C}_{ij}=\bar{C}_{ji}$. So for $N_{\rm L}$ flux bins,  we have
$N_{\rm L}(N_{\rm L}+1)/2$ independent
measurements ($\bar{C}_{ij}$ with $i\leq j$).  We also have  $N_{\rm
  L}+2$ unknowns,  given by
\be \bm\lambda=\left(C_{\rm \kappa},b_1^2C_{\rm
m},\bar{b}_{ j}{\equiv}b_{j}/b_1(j=2,\cdots,N_{\rm L}),b_1C_{\rm
m\kappa}\right) . \label{eqn:unknowns}
 \ee
 Naively speaking, when the number
of measurements is larger than the number of unknowns ($N_{\rm
L}\geq3$), in principle we can solve these equations for all the
unknowns  and extract $C_{\rm \kappa}$ rather model-independently.

However, there exists a strict degeneracy among these unknowns,
which prohibits us to solve for all of them. Specifically, Eq.
\ref{eqn:measurements} is invariant under the following
transformation:
 \ba
\label{eqn:deg}
b_{i}~~~~&\rightarrow&Ab_{i}+Bg_{i}\ , \\
\nonumber C_{\rm m}~~&\rightarrow&A^{-2}C_{\rm m}\ , \\ \nonumber
C_{\rm m\kappa}&\rightarrow&A^{-1}C_{\rm m\kappa}-A^{-2}BC_{\rm m}\ , \\
\nonumber C_{\rm \kappa}~~&\rightarrow&A^{-2}B^2C_{\rm
m}-2A^{-1}BC_{\rm m\kappa}+C_{\rm \kappa}\ . \nonumber
 \ea
Here, the parameters $A$ and $B$ are arbitrary (flux independent)
constants.

It turns out that this degeneracy arises from those cross terms
($gbC_{m\kappa}$). We can eliminate these cross terms by switching
to new variables $\tilde{b}_i$ and $\tilde{C}_\kappa$. Here,
\be
\label{eqn:hatb}
\tilde{b}_{i}\equiv \sqrt{C_{\rm m}}\left(
b_{i}+g_{i}\frac{C_{\rm m\kappa}}{C_{\rm m}}\right)\ ,
\ee and
\be
\label{eqn:hatC} \tilde{C}_\kappa\equiv C_{\kappa}(1-r^2_{\rm
m\kappa})\ ;\ r^2_{\rm m\kappa}\equiv \frac{C^2_{\rm
m\kappa}}{C_{\rm m}C_\kappa}\ .
\ee
Under these new notations,
 \be
\label{eqn:DPR}
\bar{C}_{ij}=\tilde{b}_{i}\tilde{b}_{j}+\tilde{C}_{\kappa}g_{i}g_{j}\
.
\ee
The new set of unknowns is $\bm\lambda^{\rm
new}=\left\{\tilde{C}_{\kappa},\tilde{b}_{ i}(i=1,2,\cdots,N_{\rm
L})\right\}$, and the number of these new unknowns accounts to
$N_{\rm L}+1$.

In Appendix \ref{sec:appendix1}, we mathematically prove that, when
$b$ and $g$ have different flux dependences and $N_{\rm
  L}\geq 2$, the solution of $\bm\lambda^{\rm
new}=\left\{\tilde{C}_{\kappa},\tilde{b}\right\}$ is unique.
Furthermore, for all cases we evaluated, the Fisher matrix inversion
is stable under the two conditions,  so the uniqueness of the
solution is numerically guaranteed too. Therefore, we can solve for
$\tilde{C}_{\kappa}$ and $\tilde{b}$ from the measured
$\bar{C}_{ij}$, but not $C_\kappa$ and $b$.

$\tilde{C}_\kappa$ is a biased measure of the true lensing signal
$C_\kappa$, subject to a multiplicative factor $1-r^2_{\rm
  m\kappa}$. But in practice this factor is of little importance, for
two reasons.  (1) Firstly, $r^2_{\rm m\kappa}\ll 1$ for a
sufficiently narrow redshift bin,  since the efficiency of matter in
this bin to lens a source in the same redshift bin is low. For
example, for a reasonable $\Delta z=0.2$,  $r^2_{\rm m\kappa}\sim
1\%$ at $z\sim 0.5$ and $r^2_{\rm m\kappa} \la 0.1\%$ at $z\ga 1$
(Fig. \ref{fig:one}). So the induced systematic error is of the
order $1\%$ or less, much smaller than other errors in weak lensing
measurement (Fig. \ref{fig:three}). For this reason, we can safely
neglect this error and safely treat  $\tilde{C}_\kappa=C_\kappa$.
(2) Furthermore, since by definition $r^2_{\rm m\kappa}$ is
independent of galaxy bias, $r^2_{\rm m\kappa}$ can be robustly
calculated given a cosmology and  the multiplicative correction
$1-r^2_{\rm
  m\kappa}$ can be appropriately taken into account in theoretical
interpretation. So it in principle does not cause any
systematic error in cosmological parameter  constraints.

\begin{figure}
\includegraphics[angle=270,width=84mm]{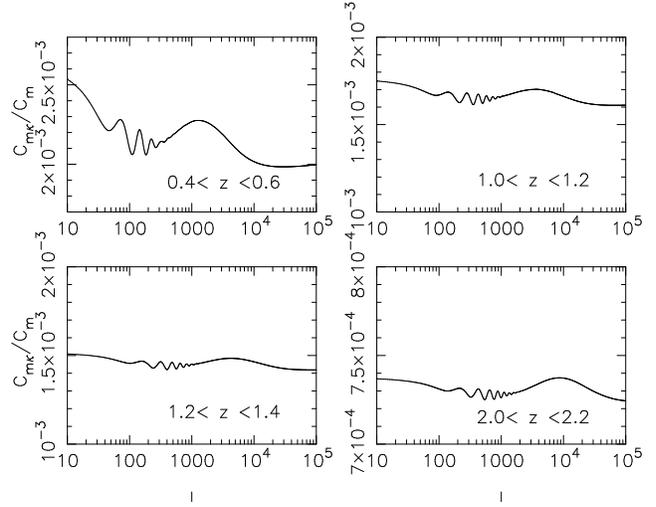}
\caption{Directly determined $\tilde{b}$ differs from the real
galaxy bias $b$, up to a scaling factor $\sqrt{C_{\rm m}}$ (flux
independent), by an additive error (flux dependent) $gC_{\rm
m\kappa}/C_{\rm m}\sim C_{\rm m\kappa}/C_{\rm m}$. Here, we plot
$C_{\rm m\kappa}/C_{\rm m}$ at different redshift bins. It decreases
with redshift and has small value of the order $\sim10^{-3}$ or less
at redshifts $z\ga0.4$. Hence the additive error $gC_{\rm
m\kappa}/C_{\rm m}$ is negligible and the flux dependence of
$\tilde{b}$ is slightly deviated from real galaxy bias $b$. So we
expect that $\tilde{b}$ offers a best determination of galaxy bias
to improve the $\kappa$ map-making by the minimal variance estimator
proposed in paper I, especially for the high redshift.}
\label{fig:two}
\end{figure}

 On the other hand, $\tilde{b}$ is a biased measure of the galaxy bias
 $b$. The prefactor $\sqrt{C_{\rm m}}$ is flux-independent. So its
 absolute value is irrelevant in the lensing map making (paper I). For
 this reason, we often neglect this prefactor where it does not cause
 confusion.  The additive error $g
 C_{\rm m\kappa}/C_{\rm m}$ is flux-dependent and indeed biases the
 $\kappa$ map-making. This effect will be quantified later in
 appendix \ref{sec:appendix3}.  Nevertheless, since the error $g
 C_{\rm m\kappa}/C_{\rm m}\sim 10^{-3}g$ (Fig. \ref{fig:two}) and since both $b$
 and $g$ are of order
 unity, to an excellent approximation this additive error is negligible and
 $\tilde{b}$ has virtually identical  flux dependence as $b$.  Hence $\tilde{b}$ offers
 an excellent template to construct the minimal variance estimator for the
 weak lensing map reconstruction in paper I. Later in \S \ref{sec:improvement} we show that
  this improved estimation of galaxy bias indeed improves the map reconstruction significantly.

This direct power spectrum determination does not rely upon priors
on galaxy bias other than it is deterministic.  In this sense, it is
robust. We now proceed to quantify its performance in galaxy
redshift surveys.

\subsection{Error sources} \label{subsec:error}

 We derive the likelihood function and adopt the
Fisher matrix analysis to estimate the errors of the direct lensing
power spectrum determination. This determination is from the
measurements of galaxy-galaxy power spectra between all pairs of
flux bins and based on the validity of Eq. \ref{eqn:measurements},
so statistical and systematic deviations from this equation will all
bias the determination.

The galaxy power spectra measured in a real survey are contaminated
by measurement noise, which is denoted as $\Delta C_{ij}$ and
throughout the paper we only consider shot noise.  On the other
hand,  our modeling of the galaxy power spectra may be imperfect,
which will cause systematic error $\delta C_{ij}$. So the real
observed power spectra  are given by \be C_{ij}=\bar{C}_{ij}+\delta
C_{ij}+\Delta C_{ij}~~~~(i\leq j)\ . \label{eqn:realmeasurements}
\ee

\subsubsection{Statistical error forecast}
Measurement error
$\Delta C_{ij}$ propagates into the weak lensing power spectrum
reconstruction and causes statistical error in the reconstructed
power spectrum. In the current paper we do not attempt to make
forecast on its cosmological constraining power. Instead we focus on
the accuracy of  the determined  power spectrum, with respect to the
true power spectrum in the given survey volume instead of with
respect to the ensemble average of the power spectrum. For this
reason,  we do not need to consider cosmic variance in the lensing
power spectrum.\footnote{It is noticed that when we compare the
determined weak lensing power spectrum with its ensemble average
predicted by theory to constrain the cosmological parameters, we
must consider the statistical error from cosmic variance. Another
point we address is that the cosmic variance still influences our
results presented in \S \ref{subsec:ska} through entering the fiducial power
spectra. Since what enters into the key equation
\ref{eqn:measurements} is not the ensemble average power spectra
from theoretical prediction, but the actual values with the right
cosmic variance in the observed cosmic volume. In the performance of
the proposed method, we neglect the cosmic variance in these real
fiducial power spectra and use the ensemble average ones instead of
them. Fortunately, this approximation is a subdominant source of
error, since the cosmic variance of each power spectrum is usually
much smaller than the ensemble average one with a large survey
volume (e.g. SKA).}

We only consider the shot noise from the poisson fluctuation of galaxy distribution as the source of
statistical error in the measurements. In this case, the likelihood function can be
well approximated by a gaussian distribution thanks to the central
limit theorem and the data covariance matrix is diagonal due to
unrelated shot noise. The Fisher matrix under this simplified
condition is \citep{Zhang2010},
\begin{equation} \label{eqn:FM}
F_{\mu\nu}=\sum_{i\le j}\frac{\bar{C}_{ij,\mu}\bar{C}_{
ij,\nu}}{\sigma^2_{ij}}\ .
\end{equation}
Throughout the paper, we use the subscript ``$\mu$'' and ``$\nu$''
to denote the unknowns ($\lambda^{\rm new}_\mu$, $\lambda^{\rm
new}_\nu$, etc.). The variance of statistical error in the observed
angular power spectrum $C_{ij}$ is \be \label{eqn:sigma2}
\sigma^2_{ij}=\langle\Delta
C^2_{ij}\rangle=\frac{1+\delta_{ij}}{(2\ell+1)\Delta \ell f_{\rm
sky}}\frac{1}{\bar{n}_{i}\bar{n}_{j}}\ . \ee Here $\bar{n}_i$ is the
average galaxy surface number density of the $i$-th flux bin.
$\delta_{ij}$ is the delta function: $\delta_{ij}=1$ when $i=j$ and
$0$ when $i\neq j$. In this paper, we adopt $\Delta \ell=0.2\ell$.
The statistical error on the parameter $\lambda^{\rm new}_{\mu}$ is
\begin{equation}
\label{eqn:systematics}
\Delta\lambda_{\rm \mu}^{\rm new}=\sqrt{(\bm F^{-1})_{\mu\mu}}\ .
\end{equation}

\subsubsection{Systematic error forecast}
\label{subsubsec:sys}

Systematic deviations from Eq. \ref{eqn:measurements} can induce
systematic errors into the reconstructed parameters,
$\delta\bm\lambda^{\rm new}\equiv\bm\lambda^{\rm re}-
\bm\lambda^{\rm tr}$. Here $\bm\lambda^{\rm re}$ is the set of
parameters to maximize the likelihood and $\bm\lambda^{\rm tr}$ is
the set of their fiducial values. The Fisher matrix can also
estimate this kind of error. We have \citep{Huterer2005,Zhang2010}
\be \delta\lambda_{\rm \mu}^{\rm new}={\bm F}_{\mu\nu}^{-1}{\bm
J_{\nu}}\ ; {\bm J}_{\nu}=\sum_{i\le j}\frac{1}{\sigma_{ij}^2}\delta
C_{ij}\frac{\partial \bar{C}_{ij}}{\partial\lambda_{\rm \nu}^{\rm
new}}\ . \label{eqn:esys} \ee Here we discuss three main sources of
systematic error.

(1) The first one arises from the galaxy stochasticity.  A
reasonable and widely adopted approximation is a deterministic bias
(no stochasticity).\footnote{There are $N_{\rm L}(N_{\rm L}-1)/2$
  independent $r_{ij}$. If we have to treat all of them as unknowns, the total
  number of unknowns will be larger than the number of independent
  measurements, for any $N_{\rm L}$. The lensing power spectrum determination would
  fail in this extreme case. Fortunately, in reality we know that
  $r_{ij}$ vanishes toward large scales. So we can carry out the determination at the limit
  $r_{ij}=0$, but with extra work to quantify the associated
  systematic bias. Nevertheless, this stochasticity problem prohibits
  the weak lensing reconstruction through cosmic magnification at
  sufficiently small scales, where the stochasticity becomes large.  } Nevertheless,
since the lensing signal is much weaker than the noise of galaxy
intrinsic clustering, we have to be careful of the stochasticity,
even if it is small. The stochasticity, at two-point statistics
level, can be completely described by the cross correlation
coefficient $r_{ij}$ between the $i$-th and $j$-th flux bins. It
biases the galaxy power spectrum modeling by \be
\label{eqn:deltaC_stochasticity} \delta C_{ij}=b_{i}b_{j}\Delta
r_{ij}C_{\rm m} ~; \Delta r_{ij}\equiv1-r_{ij}\ . \ee Plugging it
into Eq. \ref{eqn:systematics}, we obtain the induced bias in
$C_{\kappa}$. To proceed, we need a model of $r_{ij}$, which in
general depends on redshift, angular scale, flux and galaxy type.
Such modeling is beyond the scope of this work and will be postponed
until we analyze mock catalogue and observational data. For
consistency, we adopt the same toy model as in paper I: $\Delta
r_{i\neq j}=1\%$. Notice that by definition it has $\Delta
r_{ii}=0$. Readers can conveniently scale the resulting systematic
error to their favorite models  by multiplying a factor $100\Delta
r_{ij}$. This systematic error turns out to be the dominant in many
cases of the direct power spectrum determination.

\begin{figure*}
\includegraphics[angle=270,width=170mm]{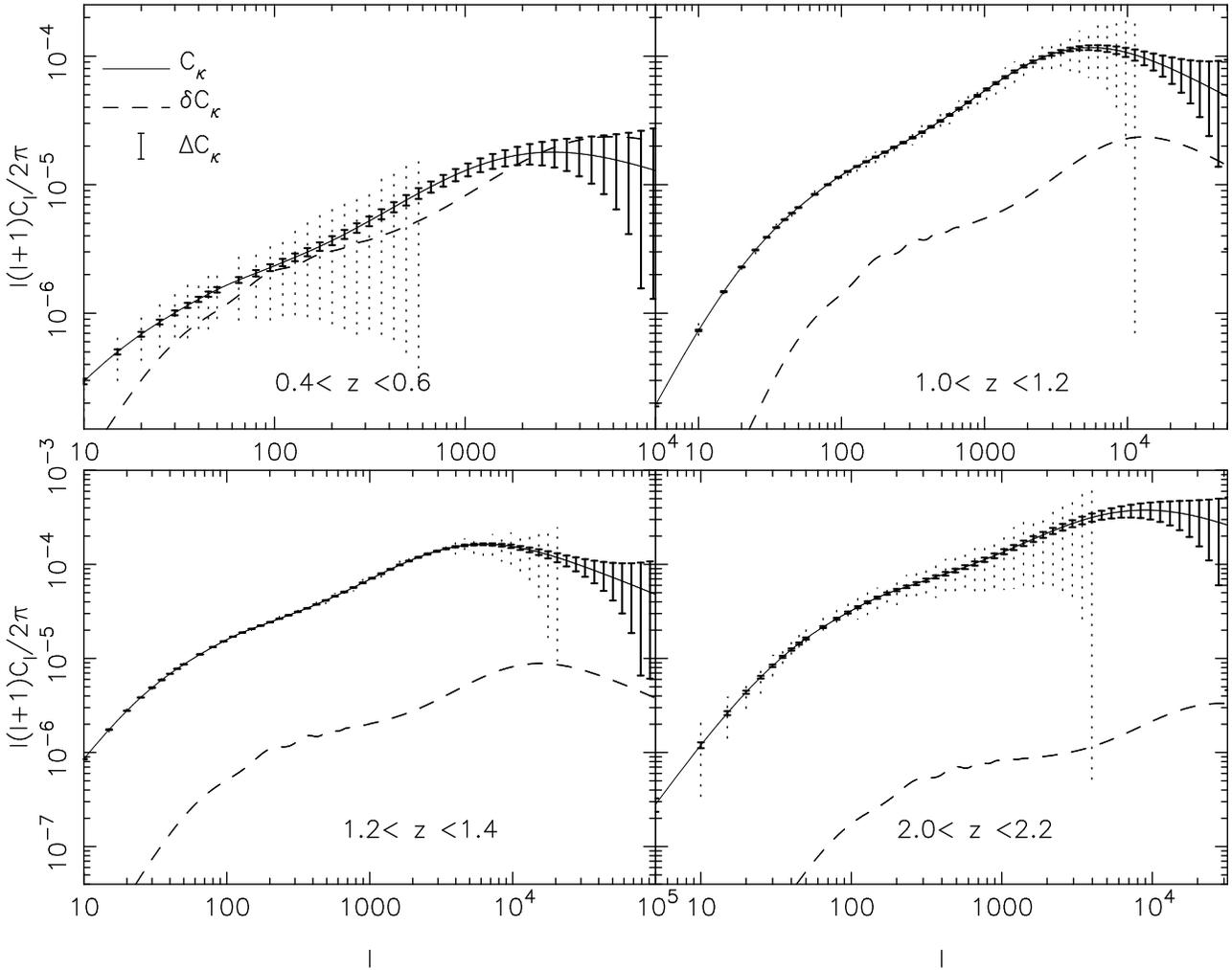}
\caption{Predicted accuracy of the direct lensing power spectrum
determination for a given redshift bin. The expected lensing auto
power spectrum $C_{\kappa}$ (solid line) increases with redshift.
Its systematic error $\delta C_\kappa$ (dashed line) from
stochasticity is dominant at low redshift (e.g. $z\sim0.5$). It
decreases with redshift and then becomes negligible at $z\sim2$. For
SKA with specifications described in paper I, the predicted
statistical error $\Delta C_\kappa$ (solid error bars) is
$5\%$-$10\%$ at low redshift $z\sim 0.5$ and $10\la\ell\la10^3$. At
redshift $z\sim1$, it can be controlled to $0.5\%$-$5\%$ level at
$10\la\ell\la10^4$. At these scales and high redshift $z\sim2$, it
is of the order $3\%$-$20\%$. For an artificial survey with the same
sky coverage as SKA but $90\%$ less galaxies, the statistical error
(dotted error bars) is still under control at redshifts $z\ga 1$.
Comparing this direct power spectrum determination with the minimal
variance $\kappa$ map reconstruction (see Fig. 6 in paper I), we
find that both reconstructions fail at $z\la 0.5$ and both lensing
power spectra can be measured with an accuracy of $\sim10\%$ at
$z\sim1$. However, the direct determination works more robustly than
the minimal variance approach at higher redshift. Especially at
$z\sim2$, the former works well and the accuracy is dominated by
statistical error, while the latter fails due to the overwhelming
systematic error from $b$-$g$ degeneracy. This error is controllable
and correctable in the direct power spectrum determination, since
this determination can reach the lower limit of errors caused by the
$b$-$g$ degeneracy.} \label{fig:three}
\end{figure*}

(2) The determination also requires precision measurement of $g$,
the prefactor of cosmic magnification. It relies on precision
measurement of the galaxy luminosity function, which could be biased
by photometry errors or errors in redshift measurement. If $g$ is
systematically biased by $\delta g$, we have \ba \delta
C_{ij}&=&\delta g_i(b_jC_{\rm m\kappa}+g_jC_{\kappa})+\delta
g_j(b_iC_{\rm
m\kappa}+g_iC_{\kappa})\\
&&+\delta g_i\delta g_j C_\kappa\ . \no \ea This $\delta C_{ij}$ is
usually much smaller than that induced by the galaxy stochasticity,
because $C_{\rm m \kappa}\ll C_m$ and $C_{\kappa}\ll C_m$. So unless
at very large scale where $\Delta r_{ij}\ll 1\%$, we can neglect the
$\delta g$ induced error. Furthermore, in this paper we will target
at SKA. It will observe billions of HI galaxies with precise
redshifts. So we believe that the luminosity function and hence $g$
can be determined to an accuracy that the induced error is
negligible.

(3) Dust extinction and photometry calibration error both bias the flux measurement and
both induce extra fluctuations in galaxy number density. The induced
fluctuation is $\propto \alpha$ instead of $\propto (\alpha-1)$,
because unlike gravitational lensing, dust
extinction and photometry error do not change the surface area.  For two reasons we do
not consider such type of errors in this paper. Firstly, for radio survey SKA, it
is free of dust extinction. Secondly, due to the different flux
dependence ($\alpha$ vs. $\alpha-1$), they can be distinguished from
the cosmic magnification. Nevertheless, we caution the readers that
weak lensing reconstruction from optical  surveys may need to  take this complexity into account.

\subsection{The performance}
\label{subsec:ska} In order to compare with the minimal variance
$\kappa$ map reconstruction  presented in paper I, we also target at
SKA to investigate the feasibility of the proposed direct lensing
power spectrum determination. Details of SKA specification are given
in paper I.

Fig. \ref{fig:three} shows  the forecasted statistical and
systematic errors in the lensing power spectrum $C_\kappa$
determination, at four redshift bins ($0.4<z<0.6$, $1.0<z<1.2$,
$1.2<z<1.4$ and $2.0<z<2.2$).  For SKA, statistical error induced by
shot noise is well under control at all the four redshift bins and
scales $\ell\la 10^4$. Even if we reduce the number density of
galaxies by a factor of $10$, corresponding to an artificial survey
with the same sky coverage as SKA but $90\%$ less galaxies, the shot
noise induced error is still insignificant at $z\sim 1$ and
$\ell\sim 10^3$.

The stochasticity induced bias is more severe.  The lensing power
spectrum increases with redshift while the
 galaxy intrinsic clustering decreases with redshift. For this
reason, the same amount of galaxy stochasticity induces larger
systematic errors at lower redshifts. Consequently, the direct power
spectrum determination fails at $z\la 0.5$ (Fig. \ref{fig:three}),
for the fiducial value of $\Delta r_{ij}=0.01$. Nevertheless, it
works well at $z\ga 1$. For $\Delta r_{ij}=0.01$, the induced
systematic error is $\sim 10\%$ at $z\sim1$ and $\sim 1\%$ at
$z\sim2$. Possibilities remain to further suppress this systematic
error. For example,  we can utilize the spectroscopic redshift
information to disregard pairs close along the line of sight,  which
are mostly responsible for this systematic error. This removal is
known to be efficient \citep{Zhang06}, and when needed, can be
applied to precision lensing reconstruction through cosmic
magnification.

\begin{figure}
\includegraphics[angle=270,width=84mm]{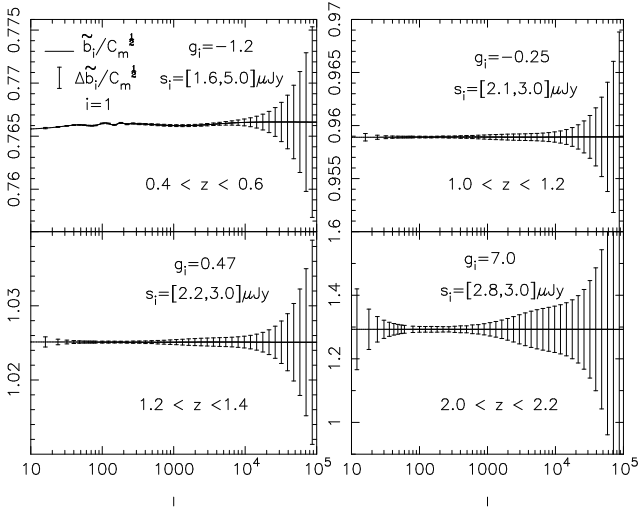}
\caption{Predicted accuracy of the direct galaxy bias $\tilde{b}$
determination. Here, we divide $\tilde{b}$ by the flux independent
scaling $\sqrt{C_{\rm m}}$ (solid line). We plot the first flux bin
($i=1$) of all four redshift bins. For a fixed redshift bin, $g$
strongly changes with flux (Fig. 3 of paper I). Here we give the
average value $g_i$ at corresponding flux bin. Since the fiducial
galaxy bias $b_i$ we adopt is scale independent (see details on
paper I) and the additive error (flux and scale dependence)
$g_iC_{\rm m\kappa}/C_{\rm m}$ is negligible, $\tilde{b}/\sqrt{C_m}$
is almost scale independent in the plot. At these four redshift
bins, systematic error $\delta \tilde{b}_i$ from galaxy
stochasticity by adopting $\Delta r_{ij}=0.01$ is below $1\%$ (this
line is too low and hence does not show up in the plot), and
statistical error $\Delta \tilde{b}_i$ (error bars) can be
controlled to better than $10\%$ at scales $10\la\ell\la10^4$. In
the plot, $\tilde{b}$ determination is getting worse with redshift.}
\label{fig:four}
\end{figure}

We make a comparison between the results of the direct lensing power
spectrum determination and those recovered from the reconstructed
$\kappa$ map in paper I (Fig. 6). Statistical error from shot noise
in Fig. \ref{fig:three} should be compared to  the minimized shot
noise in Fig. 6 of paper I.  The systematic error $\delta C_\kappa$
of Fig. \ref{fig:three} should be compared to the systematic error
$\delta C_{\rm bb}^{(2)}$ in paper I, both from the galaxy
stochasticity. At $z\la 1$, the two sets of result agree with each
other reasonably well. Due to the overwhelming error from galaxy
stochasticity, both reconstructions fails at $z\la 0.5$. At redshift
$z\sim1$, this kind of error is dominant. Nevertheless both
reconstructions can achieve an accuracy of $\sim10\%$.

However, at high redshift $z\sim 2$, the situation is different. The
direct lensing power spectrum determination works even better than
at lower redshift. To the opposite, the reconstruction presented in
paper I fails. We suspect that this failure arises from the wrong
determination of galaxy bias. In paper I, we adopted a recursive
procedure to solve the galaxy bias, with the initial guess of it \be
\left(b_i^{(1)}\right)^2=\bar{C}_{ii}/C_{\rm
m}=b_i^2+\left(g_i^2C_{\kappa}+2b_ig_iC_{\rm m\kappa}\right)/C_{\rm
m}\ . \label{eqn:binitial} \ee At low redshift, this initial guess
is nearly perfect. But at high redshift, due to increasing
$C_\kappa$ and decreasing $C_m$, the deviation from the true value
increases. Because of the degeneracy presented by Eq. \ref{eqn:deg},
the recursive procedure may not converge at the true value of galaxy
bias. The lensing map reconstructed based on the obtained false bias
is then biased. Naively we expect that the problem becomes more
severe at higher redshift.

To the opposite, the direct power spectrum determination is free of
this problem. We have proved that
$\tilde{C}_\kappa=C_\kappa(1-r^2_{\rm m\kappa})$ is what we can
solve strictly and we have numerically shown that $r^2_{\rm
m\kappa}\ll 1$ at all redshifts.

This approach also provides better determination of the galaxy bias.
We have proved that we can solve $\tilde{b}$, whose flux dependence
differs from that of $b$ only by a small additive error $ g C_{\rm
m\kappa}/C_m$. For example, this error is far below $1\%$ for all
four redshift bins plotted in Fig. \ref{fig:four}. Furthermore, it
is sub-dominant to the systematic error from galaxy stochasticity
arising from a conservative $\Delta r_{ij}=0.01$, which is under
$1\%$ at redshift up to $z\sim 2.2$. The same amount of galaxy
stochasticity induces smaller error at lower redshift. As to the
statistical error caused by shot noise, when we choose $\Delta
\ell=0.2\ell$, it is below $1\%$ at scales $10\la\ell\la10^5$ and
$0.2<z<1.6$. This statistical error increases with decreasing galaxy
number density. However even at $2.0<z<2.2$, it is controlled to
better than $25\%$ level for the bin with highest flux and hence
lowest number of galaxies. Fig. \ref{fig:four} shows the statistical
error of the first flux bin at four redshift bins. In the plot,
$\tilde{b}$ determination becomes worse with redshift. For the
highest redshift bin, the statistical error is below $10\%$ at
scales $\ell\la10^4$. This bin has quite large $g=7.0$, since we
only observe galaxies at the bright end.

Finally we want to emphasize that, to separate the galaxy intrinsic
clustering from cosmic magnification, $g$ and $b$ must have
different flux dependences. $g$ changes from positive to negative
with decreasing luminosity, but $b$ remains positive (see Fig. 3 in
paper I). So deeper surveys have advantage to measure cosmic
magnification.

\subsection{Direct determination of lensing cross power spectrum
  between different redshift bins}
\label{subsec:direct_cross} So far we focus on determining the
lensing auto power spectrum of a given redshift bin. However, a
larger portion of cosmological information is encoded in the lensing
cross power spectrum between {\it different} redshift bins. For
$N_z$ redshift bins, there are $N_z(N_z-1)/2$ cross power spectra,
but only $N_z$ auto power spectra.  So the information encoded in
these cross power spectra  is usually richer than that in the auto
power spectra. In particular, such information is essential to
perform weak lensing tomography and to measure the structure growth
rate of the universe.

We are then well motivated for a more ambitious project, namely to
determine these $N_z(N_z-1)/2$ cross power spectra. To achieve this
goal, we need not only the measurements of the galaxy clustering
within the same redshift bin ($\bar{C}_{ij}$), but also those
between different redshift bins.  We denote the redshift bins with
Greek letters ``$\alpha$'', ``$\beta$'' and flux bins with ``$i$'',
``$j$''.  We have $N_{\rm z}$ redshift bins centered at
$\bar{z}_\alpha$ ($\alpha=1,\cdots, N_z$) and  each redshift bin has
$N_{\rm
  L}$ flux bins.  So the available measurements are
$\bar{C}_{ij}^{\alpha\beta}$ ($i,j\in[1,N_L]$ and
$\alpha,\beta\in[1,N_z]$).

For the $\alpha$-th redshift bin, the available
measurements are
\ba \label{eqn:Cb} \bar{C}_{ij}^{\rm
\alpha\alpha}=b_{i}^{\rm \alpha}b_{j}^{\rm\alpha}C_{\rm mm}^{\rm
\alpha\alpha}+g_{i}^{\rm\alpha}g_{j}^{\rm\alpha}C_{\rm\kappa\kappa}^{\rm\alpha\alpha}+(b_{i}^{\rm\alpha}g_{j}^{\rm\alpha}+
b_{j}^{\rm\alpha}g_{i}^{\rm\alpha})C_{\rm m\kappa}^{\rm\alpha\alpha}\
.  \ea
We choose the independent ones with $i\leq j$.  We also  have cross correlation measurements between the $i$-th flux bin of the
$\alpha$-th redshift bin and the $j$-th flux bin of the $\beta$-th
redshift bin,
\be
\bar{C}_{ij}^{\rm\alpha\beta}=b_{i}^{\rm\alpha}g_{j}^{\rm\beta}C_{\rm
m\kappa}^{\rm\alpha\beta}+g_{i}^{\rm\alpha}g_{j}^{\rm\beta}C_{\rm\kappa\kappa}^{\rm\alpha\beta}
\ . \label{eqn:Cbf}
\ee
The above equation assumes that $\bar{z}_\alpha<\bar{z}_\beta$
($\alpha<\beta$). By requiring $\bar{z}_{\rm\beta}-\bar{z}_{\rm\alpha}\ga 0.1$, we safely neglect
a term  $\propto C_{\rm mm}^{\rm \alpha\beta}$. We still assume a
deterministic galaxy bias. $C^{\alpha\beta}_{\kappa\kappa}$ is the
lensing cross power spectrum. $b_i^\alpha C^{\alpha\beta}_{m\kappa}$
is the galaxy-galaxy lensing.  The total number of independent measurements
is
\ba
\frac{N_L(N_L+1)}{2}N_z+N_L^2\frac{N_z(N_z-1)}{2}=\frac{N_LN_z(N_LN_z+1)}{2}\ .\no
\ea
This is in general larger than the number of unknowns in
Eqs. \ref{eqn:Cb} and \ref{eqn:Cbf}.
However, a degeneracy similar to Eq.\ref{eqn:deg} exists. Eqs. \ref{eqn:Cb} and \ref{eqn:Cbf} are invariant under the
following transformation,
\ba
b_{i}^{\alpha}~~~~~&\rightarrow&Ab_{i}^{\alpha}+Bg_{i}^{\alpha}\ ,
\\ \nonumber C_{\rm mm}^{\alpha\alpha}~&\rightarrow&A^{-2}C_{\rm mm}^{\alpha\alpha}\ , \\ \nonumber
C_{\rm m\kappa}^{\alpha\alpha}~~&\rightarrow&A^{-1}C_{\rm m\kappa}^{\alpha\alpha}-A^{-2}BC_{\rm mm}^{\alpha\alpha}\ , \\
\nonumber C_{\rm
\kappa\kappa}^{\alpha\alpha}~~&\rightarrow&A^{-2}B^2C_{\rm
mm}^{\alpha\alpha}-2A^{-1}BC_{\rm m\kappa}^{\alpha\alpha}+C_{\rm
\kappa\kappa}^{\alpha\alpha}\ ,
\\ \nonumber ~~~C_{\rm
m\kappa}^{\rm\alpha\beta}~~&\rightarrow&A^{-1}C_{\rm
m\kappa}^{\rm\alpha\beta}\ ,
\\ \nonumber~~~C_{\rm \kappa\kappa}^{\rm\alpha\beta}~~&\rightarrow&C_{\rm
\kappa\kappa}^{\rm\alpha\beta}-A^{-1}BC_{\rm
m\kappa}^{\rm\alpha\beta}\ .\ea
The parameters $A$ and $B$ are
arbitrary (flux independent) constants.

Due to this degeneracy, we are not able to uniquely solve for
$C^{\alpha\beta}_{\kappa\kappa}$, the lensing cross power spectrum.
Nevertheless, following discussions in \S
\ref{subsec:reconstruction} and the appendix \ref{sec:appendix1}, we
find the solution to the following combinations is unique, \ba
\tilde{C}_{\kappa\kappa}^{\alpha\alpha}&\equiv&
C_{\kappa\kappa}^{\alpha\alpha}-\frac{(C_{\rm
m\kappa}^{\alpha\alpha})^2}{C_{\rm mm}^{\alpha\alpha}}\ , \\
\tilde{C}_{\kappa\kappa}^{\alpha\beta} &\equiv &C_{\kappa\kappa}^{\alpha\beta}-\frac{C_{\rm
m\kappa}^{\alpha\alpha}C_{\rm m\kappa}^{\alpha\beta}}{C_{\rm
mm}^{\alpha\alpha}}\no\ , \\
\tilde{C}_{\rm
m\kappa}^{\alpha\beta}&\equiv &\frac{C_{\rm
m\kappa}^{\alpha\beta}}{\sqrt{C_{\rm
mm}^{\alpha\alpha}}} \no \ ,\\
\tilde{b}_{i}^{\alpha}&\equiv& b_{i}^{\alpha}\sqrt{C_{\rm
mm}^{\alpha\alpha}}+g_{i}^{\alpha}\frac{C_{\rm
m\kappa}^{\alpha\alpha}}{\sqrt{C_{\rm mm}^{\alpha\alpha}}} \no \ .
\ea
From the appendix \ref{sec:appendix1}, we know that measurements
$C^{\alpha\alpha}_{ij}$ allow for unique determination of
$\tilde{C}^{\alpha\alpha}_{\kappa\kappa}$ and $\tilde{b}_i^\alpha$.
For $\tilde{C}_{\kappa\kappa}^{\alpha\beta}$ and $\tilde{C}_{\rm
  m\kappa}^{\alpha\beta}$, we rewrite Eq. \ref{eqn:Cbf} as
\be \bar{C}^{\alpha\beta}_{ij}=
g_j^{\beta}\left[\tilde{b}_i^\alpha\tilde{C}_{\rm
m\kappa}^{\alpha\beta}+g_i^\alpha\tilde{C}_{\kappa\kappa}^{\alpha\beta}\right]\
. \ee Since $\tilde{b}^\alpha_i$ is uniquely solved through
measurements $\bar{C}^{\alpha\alpha}_{ij}$ and $g_i^{\beta}$ is
measurable, there are only two flux-independent unknowns,
$\tilde{C}_{\rm m\kappa}^{\alpha\beta}$ and
$\tilde{C}_{\kappa\kappa}^{\alpha\beta}$. As long as $b_i/b_j\neq
g_i/g_j$, the measurements $\bar{C}^{\alpha\beta}_{ij}$
($\alpha<\beta$) uniquely determine $\tilde{C}_{\rm
m\kappa}^{\alpha\beta}$ and
$\tilde{C}_{\kappa\kappa}^{\alpha\beta}$.

$\tilde{C}_{\kappa\kappa}^{\alpha\beta}$ differs from the true
lensing cross power spectrum $C_{\kappa\kappa}^{\alpha\beta}$ by a
multiplicative factor $1-y$, \be
\tilde{C}_{\kappa\kappa}^{\alpha\beta}\equiv
C_{\kappa\kappa}^{\alpha\beta}(1-y)\ , \ee where \be \label{eqn:y}
y\equiv \frac{C_{\rm m\kappa}^{\alpha\alpha}C_{\rm
m\kappa}^{\alpha\beta}}{C_{\rm
mm}^{\alpha\alpha}C_{\kappa\kappa}^{\alpha\beta}}=\frac{r^{\alpha\alpha}_{m\kappa}r^{\alpha\beta}_{m\kappa}}{r^{\alpha\beta}_{\kappa\kappa}}\
. \ee The  three cross correlation coefficients
$r^{\alpha\alpha}_{m\kappa}$, $r^{\alpha\beta}_{\rm m\kappa}$ and
$r^{\alpha\beta}_{\kappa\kappa}$ are defined respectively as \ba
r^{\alpha\alpha}_{m\kappa}&\equiv&\frac{C^{\alpha\alpha}_{\rm
m\kappa}}{\sqrt{C^{\alpha\alpha}_{\rm
mm}C^{\alpha\alpha}_{\kappa\kappa}}}\ , \\
r^{\alpha\beta}_{\rm
m\kappa}&\equiv&\frac{C^{\alpha\beta}_{\rm
m\kappa}}{\sqrt{C^{\alpha\alpha}_{\rm
mm}C^{\beta\beta}_{\kappa\kappa}}}\ ,\no \\
r^{\alpha\beta}_{\kappa\kappa}&\equiv&\frac{C^{\alpha\beta}_{\kappa\kappa}}{\sqrt{C^{\alpha\alpha}_{\kappa\kappa}C^{\beta\beta}_{\kappa\kappa}}}\
. \no \ea
Their values are sensitive to the lensing kernel. So $y$
changes with the choice of foreground and background redshift bins.

\begin{figure}
\includegraphics[angle=270,width=84mm]{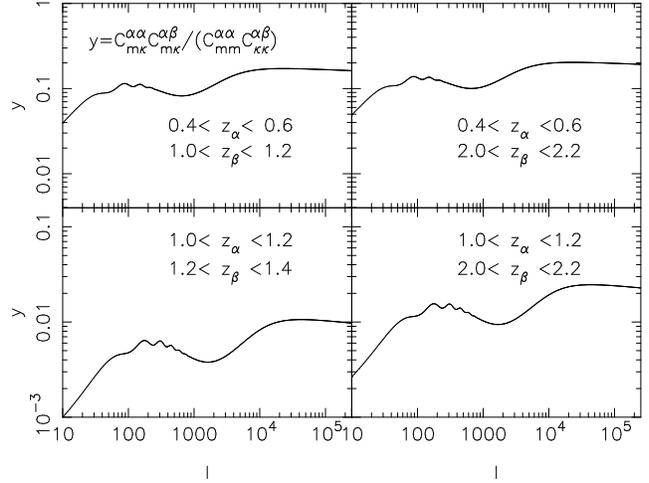}
\caption{Directly determined
$\tilde{C}_{\kappa\kappa}^{\alpha\beta}$ differs from the true
lensing cross power spectrum $C_{\kappa\kappa}^{\alpha\beta}$ by a
multiplicative factor $1-y$. Here, $y$ is determined by three cross
correlation coefficients $r^{\alpha\alpha}_{\rm m\kappa}$,
$r^{\alpha\beta}_{\rm m\kappa}$ and
$r^{\alpha\beta}_{\kappa\kappa}$. It induces error of the order
$2\%$ or less at intermediate foreground redshift $z_{\alpha}\sim
1$. At lower foreground  redshift $z_{\alpha}\sim 0.5$, the induced
error reaches up to $\sim 20\%$. Nevertheless, $y$ is correctable,
since it is free of galaxy bias and then can be calculated given a
cosmology. So the direct determination of lensing cross power
spectrum works.} \label{fig:five}
\end{figure}

By definition, $y$ is insensitive to galaxy bias and can be
calculated given a cosmology according to Eq. \ref{eqn:y}. This is a
desirable property, meaning that we can safely use
$\tilde{C}_{\kappa\kappa}^{\alpha\beta}$ to constrain cosmology
without introducing uncertainties from galaxy formation.  Fig.
\ref{fig:five} shows $y$ as a function of scales at different
foreground and background redshift bins.  It can vary from less than
$10^{-2}$ to $\sim 0.2$. So we have to take it into account when
interpreting the measured $\tilde{C}_{\kappa\kappa}^{\alpha\beta}$.

We address that $\tilde{C}^{\alpha\beta}_{{\rm m}\kappa}$ is also
free of deterministic galaxy bias.  Therefore the measured
$\tilde{C}_{\rm
  m\kappa}^{\alpha\beta}$ is also useful for cosmology. It is beyond
the scope of this paper to fully quantify the measurement accuracy of all
these quantities and to quantify cosmological information encoded in
these statistics.

Now we have demonstrated that the direct lensing power spectrum
determination indeed works. Both the lensing auto power spectra and
cross power spectra can be robustly determined from the measured
galaxy angular power spectra between different redshift and flux
bins. It hence provides a promising alternative to lensing power
spectrum measurement through cosmic shear. Even better, next section will
demonstrate that, we can improve the lensing map-making with the
improved galaxy bias determination.

\begin{figure*}
\includegraphics[angle=270,width=170mm]{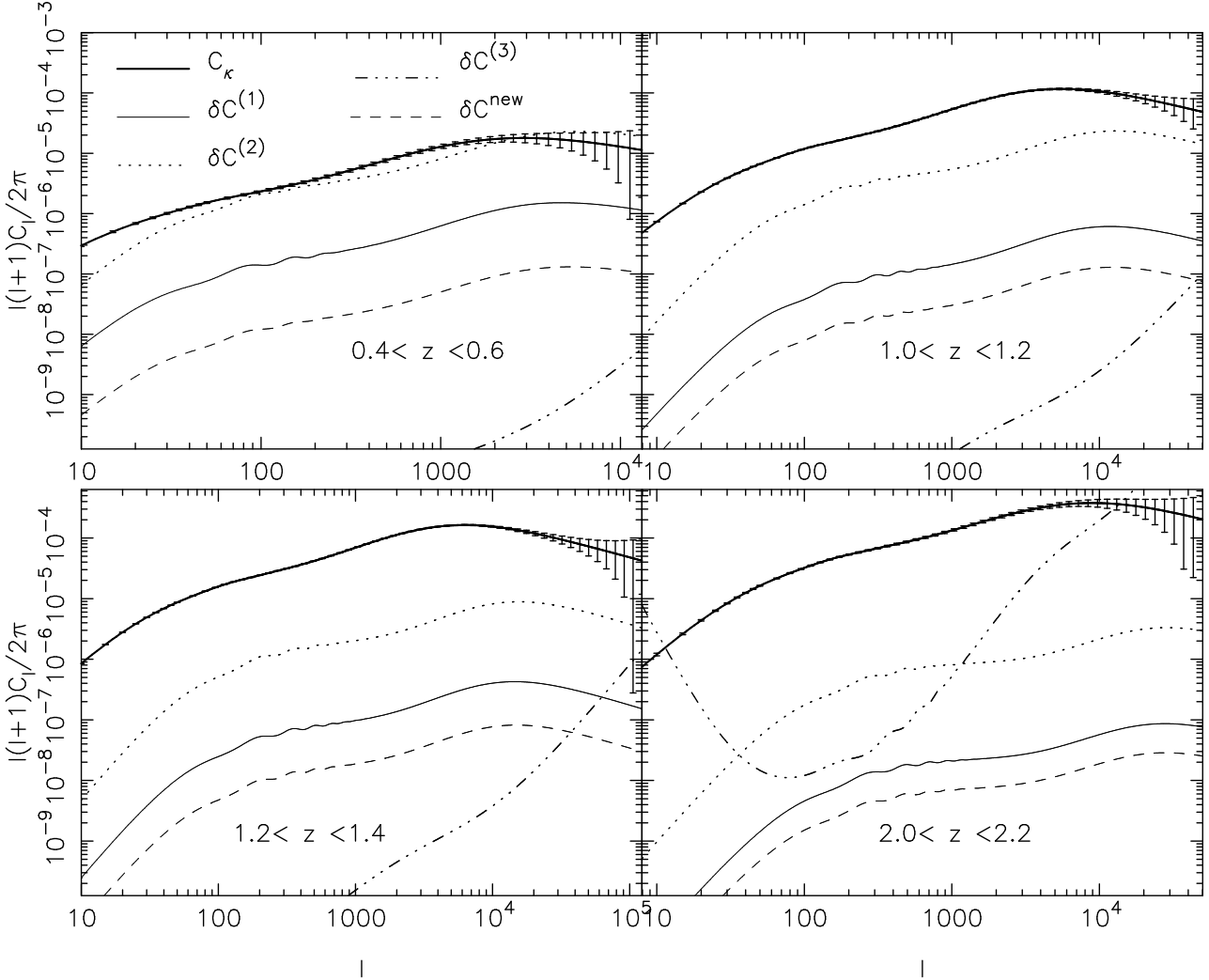}
\caption{Quality of the reconstructed $\kappa$ maps, quantified by the
  error power spectrum of each individual error sources.  Shot noise
  due to discrete galaxy distribution causes statistical errors, shown
  as error bars on top of the expected lensing auto power
  spectrum (bold solid line). We have identified four main sources of
  systematic error, whose power spectra are labeled as
  $\delta C^{(1,2,3),\rm new}$ respectively, with detailed explanation
  in appendix \ref{sec:appendix3}. The galaxy stochasticity induced error
  ($\delta C^{(2)}$, dotted line) is the most severe. It is comparable
  to the lensing signal for any pixel size (angular scale) of
  interest, at $z\sim 0.5$. The situation improves with increasing
  redshift. At $z\sim 1$,  lensing signal dominates over noises for
  pixel size of arc minute and above. Comparing with the old $\kappa$ map-making (Fig. 6 of
paper I), we find that both reconstructions fail at low redshift
$z\la0.5$, due to overwhelming $\delta C^{(2)}$. At intermediate
redshift $z\sim1$, the lensing signal overwhelms all errors and can
be measured to $\sim10\%$-$20\%$ accuracy. At $z\ga1.2$, since the
dominant error $\delta C^{(1)}_{\rm bb}$ from wrong determination of
galaxy bias in paper I is correctable in present paper, the new
measurement is robust than the old one. Especially at $z\sim2$, the
reconstruction works even better than at lower redshift on scales
$20\la\ell\la2000$, while reconstruction in paper I fails. To
conclude, the new $\kappa$ map-making supersedes the old one. One
can directly measure the lensing power spectrum from these maps,
with accuracy comparable to the direct power spectrum determination
(Fig \ref{fig:three}). This means that these maps are close to
optimal.} \label{fig:six}
\end{figure*}

\begin{figure*}
\includegraphics[angle=270,width=170mm]{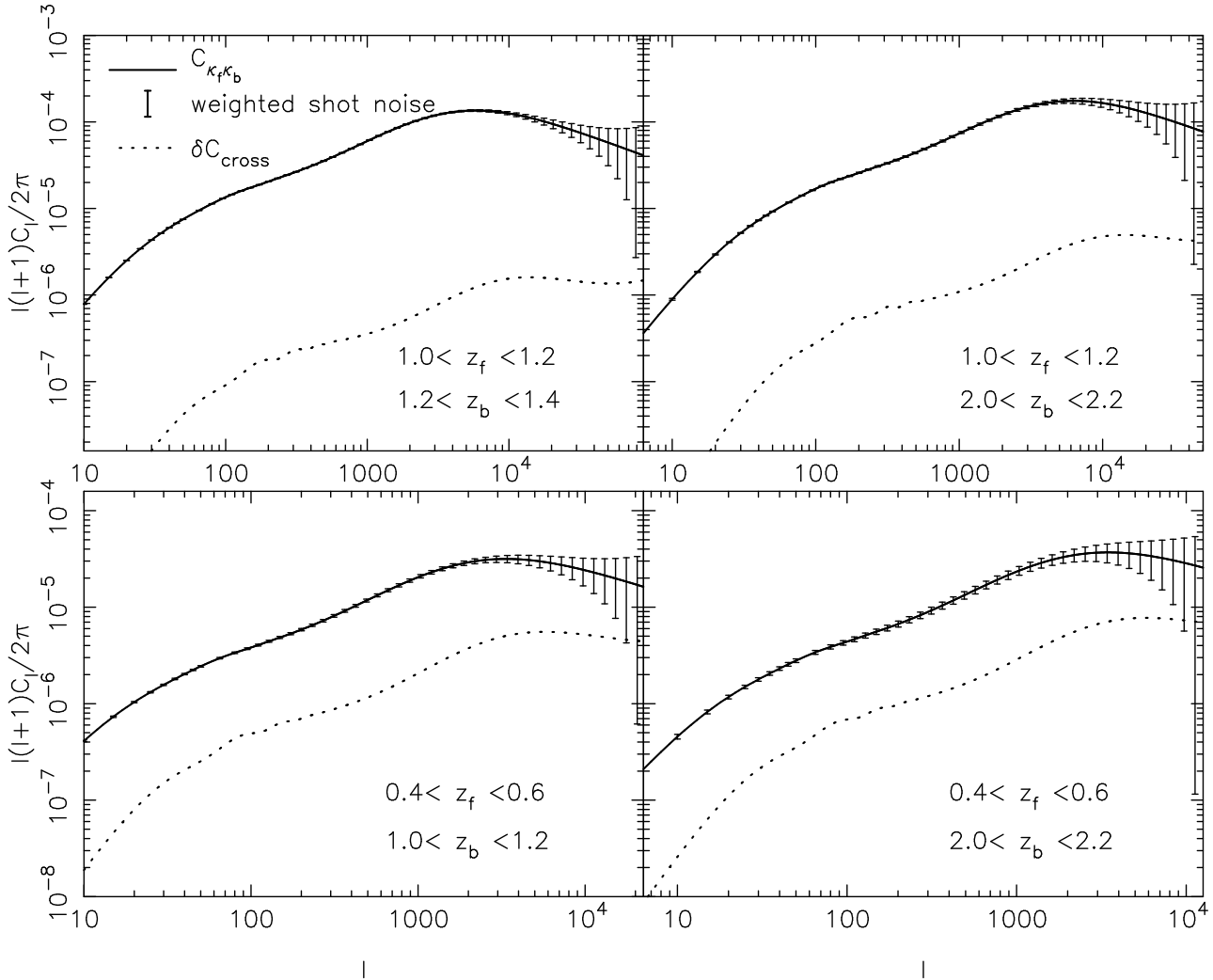}
\caption{Quality of the reconstructed $\kappa$ maps, as quantified
by the cross error power spectrum between two maps at different
redshifts. we plot the expected lensing cross power spectrum by
solid line. The error bars and dotted line show its statistical
error and systematic error, respectively. Some errors in the maps
are uncorrelated between two redshifts, such as the one arising from
the galaxy stochasticity. Systematic error arising from  the $b$-$g$
degeneracy is dominant. Comparing with the old $\kappa$
reconstruction (Fig. 7 in paper I), we find that these two
reconstructions have similar quality at low foreground redshift
(e.g. $z_{\rm f}\sim 0.5$). At higher foreground redshift (e.g.
$z_{\rm f}\sim 1$), the new maps have better quality. Hence we shall
use these improved maps to measure lensing statistics other than the
power spectrum.  } \label{fig:seven}
\end{figure*}
\section{ Improving the $\kappa$ map-making }
\label{sec:improvement}
The  minimal variance estimator of weak
lensing, proposed in paper I, requires the galaxy bias as input. It
turns out to be the limiting factor for the $\kappa$ map-making.
Fortunately, previous section shows that the galaxy bias can be
determined to high precision through direct fitting against the galaxy power
spectrum measurements. We can then improve the $\kappa$ map-making.

In paper I, we derived the unbiased minimal variance linear
estimator for the $\kappa$ reconstruction,
\be \label{eqn:kappa}
\hat{\kappa}=\sum_iw_i(\hat{b})\delta_i^{\rm L}\ ,
\ee
with $\hat{b}$ as the estimated
galaxy bias.  The subscript ``$i$'' denotes the
$i$-th flux bin of the given redshift bin. $w_i$ is the value of the weighting function $w$ at
the $i$-th flux bin.

By design, these errors in the reconstructed lensing map are additive,
\be
\hat{\kappa}=\kappa+\delta \kappa\ .
\ee
We can then define an error power spectrum $\delta C(\ell)$ for each source of error
$\delta \kappa$. It shows the contamination to the lensing signal as a
function of angular scale $\ell$. Different error sources can be
correlated, so  cross terms between different sources of error exist.
It turns out that there are many of these terms. For
brevity, we only show the error power spectrum corresponding to each
single source of error.\footnote{We caution that the error power spectra discussed here and shown in
Fig. \ref{fig:six} are to demonstrate their contaminations in the
reconstructed $\kappa$ maps. They are measures of the map quality. But they
should not be regarded as the errors in the lensing power spectrum
measured through cosmic magnification.  Correct errors should be
estimated following \S 2 (and Fig.
\ref{fig:three}).}

Furthermore, errors in maps of two redshift bins may also be
correlated. Such kind of contamination to the lensing maps can be
quantified by the cross error power spectrum $\delta C_{\rm cross}(\ell)$.

Derivations and explanations of these error sources are a little bit
technical and may not be of general interest. So we move detailed
derivations to appendix \ref{sec:appendix3}.  In Fig. \ref{fig:six},
we show four error power spectra in the reconstructed lensing maps,
denoted as $\delta C^{(1)}$, $\delta C^{(2)}$, $\delta C^{(3)}$,
$\delta C^{\rm new}$. The first three have correspondences in paper
I and represent the systematic errors caused by the $b$-$g$
degeneracy, the galaxy stochasticity and the statistical error in
galaxy bias, respectively. But the last term is new, for that we
include the galaxy cross power spectra measured between different
flux bins to determine the galaxy bias, while in paper I only the
auto power spectra of the same flux bin are utilized to infer the
galaxy bias. The target survey is SKA and target galaxies are 21cm
emitting galaxies. Survey specifications are given in paper I. Fig.
\ref{fig:seven} shows the error cross power spectrum $\delta C_{\rm
cross}$.

Fig. \ref{fig:six} shows that the map-making fails at $z\la 0.5$.
This is  due to the overwhelming error induced by stochasticity
(appendix \ref{sec:appendix3}). This is consistent with
corresponding finding in paper I.  The quality of maps improves with
increasing redshift. At $z\ga 1$, all errors are under control. This
significantly differs from the old map making method in paper I,
which degrades at $z\sim 2$. This improvement should be caused by
improved bias determination. Hence we conclude that the new
map-making method supersedes the old one in paper I.

 Some of the errors in Fig. \ref{fig:six} can be compared with that in Fig. 6 of
paper I. But detailed comparison is not necessary since there is no
exact connection between them. Nevertheless, we notice that the total
error power spectrum is comparable to that
of the direct fitting method (Fig. \ref{fig:three}) . Since in the
direct fitting we use the Fisher matrix to quantify the error and
since the error determined in this way represents the lower limit of
the true error, we conclude that the improved map-making method is
close to be optimal.

A major uncertainty in quantifying the map-making performance is the
galaxy stochasticity. As discussed in appendix \ref{sec:appendix3},
it contributes two errors, $\delta C^{(2)}$ and $\delta C^{\rm
new}$. Fig. \ref{fig:six} shows that, $\delta C^{(2)}$ not only
dominates over $\delta C^{\rm new}$ at all redshifts and all angular
scales, but also dominates over other systematic errors at almost
all redshifts/scales. As a reminder, $\delta C^{(2)}\propto \Delta
r$. We have adopted a fiducial value $\Delta r_{ij}\equiv
1-r_{ij}=0.01$.  It already forbids the lensing measurement at $z\la
0.5$. This value is reasonable at $k\sim 0.1h/$Mpc (e.g.
\citet{Bonoli2009}). But it can be much larger at smaller scales
(e.g. \citet{Bonoli2009}) and hence severely degrade the power of
weak lensing reconstruction through cosmic magnification.
Nevertheless, higher redshifts are still promising, even if the
stochasticity is a factor of 10 larger than the fiducial value
chosen in here.

Furthermore, contamination induced by galaxy stochasticity is less
severe in the determined  lensing cross power spectrum between
different redshift bins, because there is no intrinsic clustering
between two widely separately redshift bins.  Fig. \ref{fig:seven}
forecasts the error cross power spectra between different redshift
bins, which quantify correlation strength of errors in maps of two
redshifts. Among the errors,  $\delta C_{\rm cross}^{(1)}$ arising
from the $b$-$g$ degeneracy is dominant. Comparing with Fig. 7 of
paper I, we find that the new measurement has comparable accuracy at
low foreground redshift (e.g. $z_{\rm f}\sim0.5$), but considerably
higher accuracy at higher foreground redshift. We again confirm that
the new method supersedes the old one. Hence the quality of the maps
is good for cross correlation analysis.

The reconstructed lensing maps can be cross correlated with external
maps, such as lensing maps from cosmic shear or from CMB lensing
\citep{Das2011, van Engelen2012,Bleem2012, Das2013,Ade2013}. These
external correlations can have extra advantages. For example, a
major systematic error in cosmic shear is the galaxy intrinsic
alignment. But its correlation with the $\kappa$ reconstructed from
cosmic magnification of the same redshift bin is weak, due to the
vanishing lensing kernel. Major systematic error in cosmic
magnification is the residual galaxy clustering ($\epsilon
\delta_m$). But its correlation with cosmic shear is weak, again due
to the vanishing lensing kernel. So if we cross correlate the
lensing map reconstructed from cosmic magnification and the lensing
map from cosmic shear, of the same redshift bin, major systematic
errors can be significantly suppressed.

\section{Conclusions and discussions}
\label{sec:conclusions}
 In this paper we improve the lensing reconstruction through cosmic
 magnification (\citet{Zhang2005} \& paper I). It is a two step
 process.
\bi
\item Step one. For a given redshift bin, directly fitting against the
  measured galaxy cross power spectra between different flux bins to
  solve for the lensing auto power spectrum and the
  galaxy bias simultaneously. This can be extended to include the
  galaxy cross power spectra between different redshift bins to
  simultaneously solve for the lensing cross power spectrum between
  two redshifts.
\item Step two. Applying the fitted galaxy bias to the minimal
  variance estimator derived in paper I to construct the lensing
  $\kappa$ maps. These maps then can be used to measure other statistics
  such as the lensing peak abundance and lensing bispectrum, to cross
  correlate with cosmic shear or CMB lensing, etc.
\ei We have estimated its performance for the SKA survey and
demonstrated its great potential. This method has superb performance
and hence supersedes our previous works (\citet{Zhang2005} \& paper
I). Here we summarize its advantages. \bi
\item Our method differs from the traditional cosmic magnification measurement
  through foreground galaxy-background galaxy (quasar) cross correlation
  \citep{Scranton2005,Hildebrandt2009,Waerbeke2010,Menard2010,Hildebrandt2011,
Ford2012}. What the later measures is actually the galaxy-galaxy
lensing and has limited cosmological applications due to the unknown
foreground galaxy bias. To the opposite, what our method measures is
free of galaxy bias and can be used for cosmological parameter
constraints the same way as cosmic shear.
\item It is in principle applicable to all galaxy surveys with
reasonable redshift information. This is especially valuable for spectroscopic
surveys, for which cosmic shear methods do not apply.
\item It uses the extra flux information that comes for free in galaxy
  surveys.   It does not require priors on the
galaxy bias other than it is deterministic. Even better, the induced systematic
errors from stochastic bias are under control for the expected level
of galaxy stochasticity.
\item It is insensitive to dust extinction and photometry calibration error.
 This property is quite surprising given the fact that both dust
  extinction and
 photometry error affect the galaxy flux measurement.  The reason is
 that,  these two effects only alter  the galaxy flux, but do not change the
surface area as
 lensing does. For this reason, the induced galaxy density fluctuation
 $\propto \alpha$, instead of $\propto \alpha-1$ as lensing does.  In the
 appendix \ref{sec:appendix2}, we will further show that,  our approach is not only insensitive to
 random photometry errors, but also insensitive to systematic bias in
 photometry, as long as we stick to the observed $\alpha$. Dust extinction not only induces fluctuations in the galaxy
 brightness but also systematically dims the  galaxies. From the same
 argument against the photometry error, it does not induce systematic error in
 the weak lensing reconstruction through cosmic magnification.
\ei

Weak lensing reconstruction through cosmic magnification through our
method is  highly complementary to other approaches of weak lensing
reconstruction. (1) It can be used to check and control systematic
errors arising from PSF and galaxy intrinsic alignment in cosmic
shear measurement. (2)  The reconstructed lensing maps can be cross
correlated with those reconstructed from CMB lensing and 21cm
lensing to improve the lensing tomography. (3) It helps to diagnose
the impact of dust extinction in lensing reconstruction through type
Ia supernova \citep{Jonsson2010}, galaxy fundamental plane (FP)
\citep{Bertin2006,Huff2011}, the Tully-Fisher relation for late-type
galaxies \citep{Kronborg2010} and the average flux method
\citep{Schmidt2012}. All suffer from dust extinction, especially the
extinction by intergalactic gray dust, which can not be corrected
through reddening  \citep{Zhang2007}.  In contrast, our method is
insensitive to dust extinction. Comparison of the two provides a
promising way to  infer the elusive intergalactic gray dust.

Despite the great potential of weak lensing reconstruction through
cosmic magnification that we have demonstrated, there is a long list
of further studies to consolidate its role in precision cosmology.
Here we list three of them in our immediate research plan.
\bi
\item The galaxy stochasticity. We have identified
the galaxy stochasticity as the dominant source of systematic
errors. The induced systematic error can be further reduced.
Researches \citep{Tegmark1999,Bonoli2009} show that the covariance
matrix of halo clustering between different mass bins and of
different galaxy populations can be well described by the first two
principal components. If it is applied in general, it means that
only $N_L-1$ parameters, instead of $N_L(N_L-1)/2$, are required to
describe the galaxy cross correlation coefficient $r_{ij}$
($i,j=1,\cdots, N_L)$. Our method can incorporate this improved
understanding of galaxy stochasticity into account
straightforwardly. It is hence promising to solve for the lensing
power spectrum, galaxy bias and its stochasticity simultaneously.
Details will be discussed in a future paper (\citet{Yang2013}, in
preparation).
\item Tests against mock catalog. This will be done using the existing
  simulation data at Shanghai Astronomical Observatory (SHAO) ($\sim
  100 {\rm Gpc}^3$ volume in total).
\item Application to real data. As to this aspect, CFHTLS and COSMOS are
  promising targets. Both of them are sufficiently deep, with photometric
  redshift information. We emphasize that the photo-z error may be a
  main source of systematic error,  since it could bias the measurement of $\alpha(s,z)$.
  In the current paper and in paper I we target at the  spectroscopic
  survey  SKA,  so it is irrelevant.
\ei

\section*{acknowledgments}
This work is supported by the national science foundation of China
(grant No. 11025316 \& 11121062), National Basic Research Program of
China (973 Program) under grant No.2009CB24901, the CAS/SAFEA
International Partnership Program for  Creative Research Teams
(KJCX2-YW-T23) and China Postdoctoral Science Foundation
(2013M531231).

\appendix
\section{Solution to Eq. 6 is unique}
\label{sec:appendix1}
Here we prove that the solution to Eq. \ref{eqn:DPR} is unique. Suppose that there is another set of solution
$(f_1,f_2,\cdots, \sigma)$ to
Eq. \ref{eqn:DPR}, so that
\be
f_if_j+g_ig_j\sigma=\bar{C}_{ij} \ . \label{eqn:s2}
\ee
Comparing to Eq. \ref{eqn:DPR}, we have
\be
f_if_j=\tilde{b}_i\tilde{b}_j+(\tilde{C}_\kappa-\sigma)g_ig_j
\ . \label{eqn:equality}
\ee
Squaring it, we have
 \be
(f_if_j)^2=\left(\tilde{b}_i\tilde{b}_j+(\tilde{C}_\kappa-\sigma)g_ig_j\right)^2\ . \label{eqn:square}
\ee
Eq. \ref{eqn:equality} also tells us
\ba
f_i^2&=&\tilde{b}_i^2+(\tilde{C}_\kappa-\sigma)g_i^2, \nonumber \\
f_j^2&=&\tilde{b}_j^2+(\tilde{C}_\kappa-\sigma)g_j^2\ .
\label{eqn:bsquare}
\ea
Substituting  Eq. \ref{eqn:bsquare} into Eq. \ref{eqn:square}, we have
\be
(\tilde{C}_\kappa-\sigma)(\tilde{b}_ig_j-\tilde{b}_jg_i)^2=0
\ . \label{eqn:uniqueness}
\ee
Plugging the $\tilde{b}$-$b$ relation (Eq. \ref{eqn:hatb}),  we obtain
\be
(\tilde{C}_\kappa-\sigma)(b_ig_j-b_jg_i)^2=0\ .
\ee
In general, $b$ and $g$ have different dependences on flux and hence
$b_i/b_j\neq g_i/g_j$. For example, $g$ for faint galaxies can be
negative while $b$ remains positive. We then obtain $\sigma=\tilde{C}_\kappa$and $f_i=\tilde{b}_i$.

Hence we prove that, the solution to Eq. \ref{eqn:DPR} is unique, as
long as that the number of flux bin $N_L\geq 2$ and that the galaxy
bias and $g$ have different flux dependences. In the main text, we
have shown that the solution
$\tilde{C}_\kappa$ only differs from the true lensing power spectrum
$C_\kappa$ by a negligible multiplicative bias. We  can then draw the
conclusion that by directly fitting the measured galaxy clustering
between different flux bins (but of the same redshift bin), the
lensing auto power spectrum can be determined uniquely.

\section{Photometry error and dust extinction} \label{sec:appendix2}
Photometry error operates onto the lensing magnified flux. For a
galaxy with intrinsic flux $s$ and given magnification $\mu$, the
observed flux $s^{\rm O}$ is \be s^{\rm O}=s\mu\times (1+p)\ . \ee
Here we denote $p$ as the photometry error. It may have a nonzero
mean ($\langle p\rangle\neq 0$) and non-negligible fluctuations
around the mean. Through the galaxy number conservation, we have \be
n(s)dsdA=n^{\rm O}(s^{\rm O})ds^{\rm O}d(A\mu)\ . \ee Here $A$ is
the surface area, which is amplified by lensing by a factor $\mu$.
We then have the relation between the observed galaxy distribution
$n^{\rm O}$ and the intrinsic distribution $n$, \be n^{\rm O}(s^{\rm
O})=\frac{1}{\mu^2(1+p)}n\left(\frac{s^{\rm O}}{\mu(1+p)}\right)\ .
\ee Here, we have neglected the flux dependence in $p$. This
approximation is reasonable if the photometry error does not
strongly depend on the flux.

We can Taylor expand $n$ around $s^{\rm O}$ to linear order in $p$
and $\mu-1\simeq 2\kappa$. In this way, the coefficients in front of
$p$ and $\kappa$ are functions of $n$. However, since $\langle
p\rangle\neq 0$, even to  the first order approximation, $\langle
n^{\rm O}\rangle\neq \langle n\rangle$. This means that we can not
directly calculate these coefficients from observables (e.g. $n^{\rm
O}$).

We circumvent this problem by  defining another galaxy flux
distribution function $n^{\rm P}$, given by $n(s)ds=n^{\rm P}(s^{\rm
P})ds^{\rm P}$ in which $s^{\rm P}\equiv s(1+p)$. We then have \be
n^{\rm O}(s^{\rm O})=\frac{1}{\mu^2}n^{\rm P}\left(\frac{s^{\rm
O}}{\mu}\right)\simeq n^{\rm P}(s^{\rm O})(1+g^{\rm P}\kappa)\ . \ee
Here, $g^{\rm P}\equiv 2(-d\ln n^{\rm P}/d\ln s^{\rm P}|_{s^{\rm
O}}-2)$.

It is now clear that the cosmic magnification expression is still
applicable, as long as we replace the intrinsic galaxy distribution
$n$ with $n^{\rm P}$ and replace $g$ with $g^{\rm P}$. Furthermore,
since we have $\langle \mu\rangle=1$ and $\langle
(\mu-1)^2\rangle=O(10^{-3})$, $\langle n^{\rm O}(s^{\rm
O})\rangle=\langle n^{\rm P}(s^{\rm O})\rangle$ is a good
approximation. Under this limit, $g^{\rm P}\simeq 2(-d\ln n^{\rm
O}/d\ln s^{\rm O}|_{s^{\rm O}}-2)$, an observable.

Hence  photometry error does not bias the magnification coefficient
($g^{\rm P}$) and in this sense does not bias cosmic magnification
measurement. But it indeed introduces new fluctuations  in the
galaxy density distribution.  Taylor expanding the relation $n^{\rm
P}(s^{\rm O})=n(s^{\rm O}/(1+p))/(1+p)$, we obtain \be n^{\rm
O}(s^{\rm O})\simeq \bar{n}(s^{\rm O})\left(1+\delta_{\rm g}+g^{\rm
P}\kappa+\big(1+\frac{g}{2}\big)p\right)\ , \ee or equivalently, \be
\delta_g^{\rm O}\simeq \delta_{\rm g}+g^{\rm
P}\kappa+\big(1+\frac{g}{2}\big)\big(p-\langle p\rangle\big)\ . \ee
Notice that $g\neq g^P$ with the presence of photometry error. Since
the photometry error $p$ is a random number, it causes fluctuation
in galaxy distribution. This fluctuation can be distinguished from
the cosmic magnification in two ways. Firstly, they have different
flux dependences ($1+g/2$ vs. $g^{\rm P}$). Secondly, they have
different spatial clustering. $p-\langle p\rangle$ may resemble a
shot noise like spatial clustering, although its amplitude can vary
across the sky due to spatial variation in photometry calibration
accuracy.

Since we allow $\langle p\rangle\neq 0$ and allow $p$ to fluctuate
across the sky, the above discussion also applies to dust
extinction. Then we conclude that both photometry error and dust
extinction do not bias our weak lensing reconstruction through cosmic
magnification.

\section{Error sources in the reconstructed lensing maps}
\label{sec:appendix3}
A number of errors in the map making comes as follows. In the limit of deterministic bias
($\delta_i=b_i\delta_m$),  we have\footnote{There are errors which can not
be described by $\epsilon$.}
\be \hat{\kappa}\rightarrow
\kappa+\epsilon \delta_m \ ,
\ee
 where
\be \epsilon\equiv \sum_i
w_i(\hat{b})b_i \ .
\ee
We require $\sum_i w_i(b)b_i=0$ in order to eliminate
the galaxy intrinsic clustering. $w$ satisfying this
condition while minimizing the rms error in the map-making is derived
in paper I. This optimal weighting function has an analytical
expression. It is  uniquely  fixed by the galaxy
luminosity function and the galaxy bias $b$. Notice that  $w_i$
depends not only on the bias at the $i$-th flux bin, but also biases
at other bins.  We highlight this dependence by explicitly showing
$b$, instead of $b_i$,  as the argument of $w_i$.

$w_i$ is invariant under a flux independent scaling in the galaxy
bias b (paper I). In previous sections we show that we can determine
$\tilde{b}_i\equiv \sqrt{C_{\rm m}}(b_i+g_i C_{\rm m\kappa}/C_{\rm
m}$) to high accuracy. So in deriving the optimal estimator $w$, we
do not need to worry about the absolute value of $C_{\rm m}$, which
is flux-independent. Hereafter we will ignore this prefactor.

By reconstruction, our estimator guarantees $\sum_i
w_i(\hat{b})\hat{b}_i=0$, but not the desired $\sum_i
w_i(\hat{b})b_i=0$ . Errors in $\hat{b}$ ($\hat{b}\neq b$) cause
$\epsilon\neq 0$ and induce additive errors in the reconstructed
lensing maps. Taylor expanding $w(\hat{b})$ around the true value
$b$, we have \ba \label{eqn:epsilon}
\epsilon&=&\sum_{ij}\left[\frac{\partial w_i}{\partial
    b_j}(\hat{b}_j-b_j)b_i\right]\\
&+&\frac{1}{2}\sum_{ijk}\left[\frac{\partial^2w_i}{\partial
    b_j\partial b_k}(\hat{b}_j-b_j)(\hat{b}_k-b_k)b_i\right]+\cdots\no \ .
\ea A systematic error has $\langle \hat{b}-b\rangle \neq 0$. So we
just keep the linear term above to evaluate $\epsilon$. A
statistical error has  $\langle \hat{b}-b\rangle=0$, but  $\langle
(\hat{b}-b)^2\rangle\neq 0$. Since $w(b)$ is nonlinear in terms of
$b$, $\langle \epsilon\rangle\neq 0$. So even a statistical error in
$b$ can be  rendered into systematic error  in the $\kappa$
reconstruction.

We then have
\be \delta C=\epsilon^2
C_{\rm m}+2\epsilon C_{{\rm
    m}\kappa}\label{eqn:dC} .
\ee Notice that although usually $\epsilon\ll 1$, $\epsilon^2 C_{\rm
m}$ is not necessarily smaller than $2\epsilon C_{{\rm m}\kappa}$,
because $C_{\rm m}\gg C_{{\rm m}\kappa}$ by one or two magnitude
(see Fig. 1 of paper I).

We also have
\be \delta C_{\rm cross}=\epsilon_f C_{\rm
m_f\kappa_b}\ . \ee Here, following notations in paper I, we use the
superscript ``$b$'' to denote the background (higher redshift) bin
and the superscript ``$f$'' to denote the foreground (lower
redshift) bin. In the expression, we have neglected the correlation
$C_{\rm m_fm_b}$ between foreground and background matter
distributions. It is natural for non-adjacent redshift bins with
separation $\Delta z\ga 0.1$, since foreground and background
galaxies have no intrinsic correlation. For two adjacent redshift
bins (e.g. the left-upper panel in Fig. \ref{fig:seven}), there is
indeed a non-vanishing matter correlation. However, this correlation
is also safely neglected since  both the foreground and background
intrinsic clustering are sharply suppressed by factors
$1/\epsilon_{\rm f,b}$, respectively. For this reason, stochasticity
no longer causes a term like $\delta C^{(2)}$ (discussed later).

For the cross  power spectrum between different bins, the error
power spectrum shown in Fig. \ref{fig:seven} is the errors in the
lensing cross power spectrum measured through cosmic magnification.
Since the cross terms between different sources of error no longer
exist in the cross power spectrum measurement.

\subsection{Systematic error caused by the $b$-$g$ degeneracy}
In the ideal case of no other sources of error, the $b$-$g$
degeneracy (Eq. \ref{eqn:deg}) still causes a systematic error in
the determined galaxy bias $\hat{b}_i=\tilde{b}_i\propto b_i+g_i
C_{m\kappa}/C_m\neq b_i$. As we discussed earlier and as in paper I,
flux-independent scaling in the galaxy bias (e.g. $\sqrt{C_m}$) does
not affect the map-making. So we will ignore the $\sqrt{C_m}$
prefactor in $\hat{b}$.  Following the notation in paper I, we
denote such error with superscript
 ``(1)'',
 \be
\label{eqn:e1}
\epsilon^{(1)}\simeq \frac{C_{\rm
m\kappa}}{C_{\rm m}}\left[\sum_{ij}\frac{\partial w_i}{\partial
    b_j}\bigg|_{b_j}g_jb_i\right]\ .
\ee
 Since $\epsilon^{(1)}\sim C_{{\rm m}\kappa}/C_{\rm
m}=O(10^{-3})\ll 1$ (Fig. \ref{fig:two}),  the galaxy intrinsic
clustering is heavily suppressed.

\subsection{Systematic errors induced by galaxy stochasticity}
Stochasticity biases the $\kappa$ reconstruction in two ways. The
first has been identified in paper I (Eq. 26). Even if we correctly
figure out the deterministic component of galaxy bias, stochasticity
does not allow us to completely remove the intrinsic galaxy
clustering. The residual part is
\be
\delta C^{(2)}=-\left[
\sum_{ij} w_i({b})w_j({b})b_ib_j\Delta
  r_{ij}\right]C_m \label{eqn:C2} .
\ee Following the notation in paper I, we denote this error with a
superscript ``(2)''. This error does not affect the cross
correlation measurement.

The galaxy stochasticity also causes systematic bias in the
determined galaxy bias (Eq. \ref{eqn:esys}) and hence biases the
$\kappa$ reconstruction through $w(b)$. It arises since we include
the cross power spectra between different flux bins to measure
galaxy bias. This one does not have counterpart in paper I, where
only the auto power spectra  of the same flux bin are utilized to
infer the galaxy bias. We will denote this new type of error with a
superscript ``new''.  The corresponding $\epsilon^{\rm new}$ can be
calculated with Eqs. \ref{eqn:esys}, \ref{eqn:deltaC_stochasticity}
\& \ref{eqn:epsilon}, given by \be \epsilon^{\rm
new}\simeq\sum_{ij}\frac{\partial w_i}{\partial
    \tilde{b}_j}\bigg|_{\tilde{b}_j}\delta\tilde{b}_jb_i\ .
\ee
\subsection{Systematic error caused by statistical error in galaxy bias}
As discussed earlier, statistical error in galaxy bias also induces
systematic error in the $\kappa$ reconstruction. The induced
systematic error in $\hat{\kappa}$ is \be \epsilon^{(3)}\simeq
\sum_{ij}\frac{\partial w_i}{\partial
    \tilde{b}_j}\bigg|_{\tilde{b}_j}\Delta\tilde{b}_jb_i+\sum_{ijk}\left[\frac{\partial^2w_i}{\partial
    \tilde{b}_j\partial \tilde{b}_k}\bigg|_{\tilde{b}_j}{\bf
    B}^{-1}_{jk}\tilde{b}_i\right]\ .
\ee Here $\bf B^{-1}$ is a sub-matrix of the Fisher matrix $\bf
F^{-1}$ corresponding to parameters of galaxy bias. Although the
ensemble average of the first term is zero, this term does
contribute to $[\epsilon^{(3)}]^2$ in Eq. \ref{eqn:dC}. This error
has a counterpart in paper I. Following  the notation there, we
denote it with a superscript ``(3)''.

\end{document}